\documentclass[useAMS,usegraphicx,usenatbib]{mn2e}
\usepackage{threeparttable}
\usepackage{times}
\usepackage{epsfig}
\usepackage{textcomp}
\usepackage{amssymb}

\def\xmm{{\it XMM-Newton}}

\title[Rms variability properties of the iron K$\alpha$ line in Seyfert galaxies.]{Rms variability properties of iron lines in Seyferts.}

\author[S. Bhayani et~al.]{Shyam Bhayani$\thanks{shyam.bhayani03@imperial.ac.uk}$ and Kirpal Nandra \\
Astrophysics Group, Imperial College London, Blackett Laboratory, Prince Consort Road, London SW7~2AW\\}

\begin{document}


\date{Accepted. Received}

\pagerange{\pageref{firstpage}--\pageref{lastpage}} \pubyear{2009}

\maketitle

\label{firstpage}

\begin{abstract}
We present an analysis of the rms variability spectra of a sample of 18 observations of 14 Seyfert galaxies observed by \xmm, which exhibit sufficient variability and signal-to-noise ratio to examine the variations in the iron K-band. The narrow core of the K$\alpha$ line at 6.4 keV, seen universally in Seyferts, shows minimal evidence for variability and is always less variable than the continuum, supporting an origin in distant material such as the torus. At least half the observations do show evidence for variations in the wider iron K-band, however, and in at least 5 cases the excess line variations appear to be broad. The simplest prediction -- that the broad emission line is as variable as the continuum -- is generally not confirmed as only two observations show this type of behaviour. In four cases, the red wing of the line is more variable than the power-law continuum and extends down to energies of $\sim$ 5 keV. Three observations show strong variability blueward of the line core that could also be from the disk, but alternatively might be due to emission or absorption by other hot or photoionised gas close to the nucleus. In cases where this excess blue variability is present, it is not always seen in the time-averaged spectrum. Six observations possess a broad iron line in the time-averaged spectra but with an invariant red wing, and three of these six show no variability across the entire iron line region. This suggests a decoupling of the continuum and reflection component, perhaps due to light bending or other anisotropic effects as has been suggested for MCG-6-30-15 and other narrow-line Seyfert 1s. Four objects in our sample have two observations with well-defined rms spectra, and all are seen to change their iron line variability properties when re-observed. This, and the great diversity of iron line variability properties indicated by the rms spectra makes it extremely difficult to interpret the variability in any simple framework. However, a key result is that the rms spectra of objects such as NGC 3516 do not agree with complex absorption effects mimicking the broad red wing. Instead, the implication is that the central regions of AGN close to the black hole are extremely complex, with the central accretion flow being chaotic and highly variable. Anisotropy and strong gravitational effects are also most likely in effect.
\end{abstract}


\section{INTRODUCTION}

The spectra of many Seyfert galaxies have been found to contain an excess at around 6.4 keV due to iron K$\alpha$ emission (Pounds et al. 1990; Nandra \& Pounds 1994). This line can be used as a tool to probe the inner regions of black holes at the centre of active galactic nuclei (AGN) as it is broadened and asymmetric, which is thought to be due to Doppler motions and relativistic effects close to the central black hole (Fabian et al. 1989). Such broadened emission around the iron line is found in a substantial fraction of AGN (Nandra et al. 1997b). A recent survey employing \xmm\ required broadened K band emission in two thirds of a sample of Seyferts (Nandra et al. 2007, hereafter N07). The high signal-to-noise ratio of \xmm\ spectra provides an opportunity to understand the variability of the iron line through rms spectra (e.g. Vaughan et al. 2003). This is largely model-independent and so it can complement the time-averaged spectral analysis and discriminate between models. For example, because the broad iron K$\alpha$ line is excited by the continuum, and arises close to the central black hole, one would expect that objects with a broad line would show significant excess variability in their rms spectra compared to the pure continuum, or alternatively, a featureless spectrum around the iron line when the rms is divided by the mean. Note that in the current work, we use the rms spectrum without normalizing to the mean, so when discussing previous work using the normalized rms spectra we say so explicitly. 

The rms variability spectra of relatively few objects has so far been determined, and most studies have focused on the continuum variations, rather than the iron K$\alpha$ emission line. A notable exception is MCG-6-30-15 (Fabian et al. 2002). This shows a narrow dip in the normalised rms spectrum around 6.4 keV, which could be associated with the narrow iron line component arising in the outer regions of the disk, or the torus envisaged in AGN unification models (Krolik \& Kallman 1987). Emission this far from the black hole accounts for the dip in the normalised rms as it should be much less variable than the continuum. A narrow, neutral iron K$\alpha$ component appears to be ubiquitous in Seyfert spectra (Yaqoob \& Padmanabhan 2004; N07), so a signature of non-variability at 6.4 keV could well be present in the variability spectra of most objects. Turning to the broader emission, in an early \xmm\ observation Ponti et al. (2004) found an excess of variability between 4.7-5.8 keV, which is most likely due to the broad, redshifted iron line component emitted from a disk around a spinning black hole. Alternatively, Ponti et al. suggest it could represent absorption of the power law continuum by iron in material infalling into the black hole. Regardless of the cause of the excess in the normalised rms, it tells us that the broad component is more variable than the continuum. In contrast, a later and longer \xmm\ observation showed a dearth of variability around the iron K$\alpha$ line (Vaughan \& Fabian 2004). Miniutti et al. (2007) have studied the normalised rms spectrum of MCG-6-30-15 across a wider energy range with the broad-band instruments aboard {\it Suzaku} and found that the normalised rms decreased with energy above 3.0 keV and there is a drop in variability at 6.4 keV. The overall decrease is interpreted as being due to the reflection component being less variable than the continuum. Normalised rms spectra and other techniques such as time-resolved and difference spectra have been used to suggest MCG-6-30-15 as a strong case for the two-component model where a variable power-law with a constant slope is superimposed on a hard reflection component that is nearly constant (McHardy et al. 1999; Shih et al. 2002; Taylor et al. 2003; Fabian \& Vaughan 2003). This is contrary to the predictions of the simplest accretion disk model, where it is expected that variations in the reflection spectrum follow those in the power-law continuum. This would result in a flat normalised rms spectrum with no dip or excess in the iron line band. An absorption origin has also been suggested to produce an apparently broad red wing in AGN (Reeves et al. 2004), including in MCG-6-30-15 (Miller et al. 2008). In this case, the variations in  the red wing should then still match the time-averaged red wing as the red wing has been formed by the continuum passing through an absorber and thus the two components should possess the same variability. 

The variability spectrum for MCG-6-30-15 has helped in understanding the origin of the iron line present in the time-averaged spectrum. The aim of this paper is to investigate the rms spectra of a large sample of Seyfert galaxies around the iron line, to shed similar light on the iron K$\alpha$ line variability of AGN in general.

\section{OBSERVATIONS AND ANALYSIS}

The parent sample of objects used in the current study is essentially the same as that of N07, and the data and reduction methods used here are identical to that work. Briefly, N07 only Seyferts at redshift less than 0.05 with observations available in the \xmm\ public archive as of 1 Jan 2006. Seyfert 2s are excluded due to obscuration of the central object by the torus. The observations were recorded with the EPIC-pn CCD camera and details such as the data, exposure, mode and filter are given in Table 2 of N07. An observation was included in the sample if there was a minimum of 30,000 counts in the 2-10 keV spectrum after the data were screened for periods of high background. The final group in N07 contained 26 objects with a total of 36 observations. The main difference in the current work is that we added a third, long (303ks) observation of Mrk 766 from May 2005 (OBSID 0304030101) to the sample.

\subsection{Variability Spectral Analysis \label{sec:vsa}}

Here we define the rms as in Nandra et al. 1997a:

\begin{equation}
\sigma_{rms}=\sqrt{\frac{1}{N}\displaystyle\sum_{i=1}^N [(X_{i}-\mu)^{2}-\sigma_{i}^{2}]}
\label{eq:evar}
\end{equation}

where $\it{N}$ is the number of time bins in the light curve, $X_{i}$ and $\sigma_{i}$ are the counting rates and uncertainties, respectively, in each bin and $\mu$ is the unweighted arithmetic mean of the counting rates. This equation subtracts the variance due to measurement errors from the total variance. The resulting spectrum will still contain features due to the instrumental response, but these can be accounted for using the same effective area and response matrix employed in the time averaged spectra (N07). This also allows for a direct, quantitative comparison between the rms and time-averaged spectra. When searching for line variations, this makes the unnormalised rms spectrum easier to interpret and arguably more sensitive than analysis of the normalised version.

The energy range is restricted to 2.5 - 10 keV to minimize the effects of photoelectric absorption and/or soft excess component, which are manifest primarily at softer energies. This is justifiable as we are primarily interested in the iron line and its relationship with the continuum. As defined above, It is possible for the excess variance, $\sigma_{rms}^{2}$, to be negative if, due to statistical fluctuations, the intrinsic variability is small and outweighed by the measurement errors. The rms cannot then be defined and if this occurs even in a single bin the observation cannot be used. The major consequence of this problem is in the choice of binning the rms spectra. With small bins, many observations will be rejected due to negative rms values, reducing the sample size. On the other hand, if the bins become too large, there will be insufficient resolution to distinguish the line variations from those of the continuum. To strike a balance between the resolution of the spectra around the iron line a sample size, we adopted the same set of 11 energy bins between 2.5-10 keV for each spectrum, with 400 eV resolution in the iron band (4.5-7.5 keV). This resolution is sufficient to examine the variability of the broad and narrow components of the line, and yields fully-defined rms spectra for a total of 18 observations of 14 objects. The rms spectra were calculated using time bins of 1000s.

The uncertainties in the rms measurements for each band were calculated using Monte Carlo simulated light curves. A Gaussian distribution was devised for each time bin with mean and standard deviation set to the observed counting rate and uncertainty respectively in that bin. The counting rate for each simulated light curve bin was found by drawing a random deviate from the bin's distribution. 10,000 synthetic light curves were simulated and the rms spectrum for each was calculated. Finally, the $1\sigma$ uncertainty of each bin was set to the standard deviation of the 10,000 rms values for that bin.

The rms spectra broadly follow a power law form, as do the time-averaged spectra. The variability around the iron line region is thus easier to study by taking the ratio of variability to a power-law. A power-law fit to the 2.5-10 keV interval was obtained, excluding the iron line band in the 4.5 - 7.5 keV region. Galactic absorption was included using the XSPEC $\it{wabs}$ model with $N_{\rm H}$ values given in N07. N07 found that some of the observations in our sample show additional effects of ionized absorption local to the AGN. We accounted for this in cases where it was required using the XSTAR table model $\it{cwa18}$ described in N07, with parameters fixed to those found by N07. The power-law index and normalization were left free to vary. The data/model ratios are shown in Fig. 1 where the ratio of the time-averaged spectrum to a power-law model excluding 4.5-7.5 keV has also been overplotted.

\subsection{Variability significance \label{sec:vs}}

An excess in the iron line region over the fitted continuum implies iron line variability, and the errors shown in Fig. 1 provide a crude visual impression of the significance of those variations. We have also made a more explicit estimate of the significance of any line variability by performing additional simulations of the expected rms spectrum of a power-law continuum with no emission line components. A continuum light curve was simulated by using the $\it{fakeit}$ function in XSPEC to create a power-law spectrum for each 1000s time bin. The model is the same as the continuum used to fit the real rms spectrum (i.e. those shown in Fig. 1), but the normalization of the power-law for each bin was set such that the total count rate of the four continuum bins (i.e. those between 2.5-4.6 keV and 7.4-10.0 keV), matched the observed count rate in this energy range. The 4.6-7.4 keV interval is excluded in the renormalization, as this will contain any contribution by line components, whereas the rest of the energy range should contain only the continuum. These simulations therefore yield fake light curves in each energy bin, based solely on the observed continuum spectrum and light curve, with random Poisson noise. These simulations were repeated 1000 times and the excess variance spectrum was calculated for each one, just as it was for the real light curve. We can then compare how many times the excess variance in the fake continuum rms spectrum exceeds that of the observed rms spectrum, which contains the emission line. This then yields the probability that the observed variance exceeds the continuum model by chance, and hence the significance of the line variations. We deem an excess in an individual to be significant if its confidence level exceeds 95\%. In Table 1, we list all all bins with a confidence level of at least 80\%, as several adjacent bins with moderately high confidence might indicate a broad variability excess, even if individually they do not all exceed 95\% confidence.

\begin{figure*}
\includegraphics[angle=0,width=58mm]{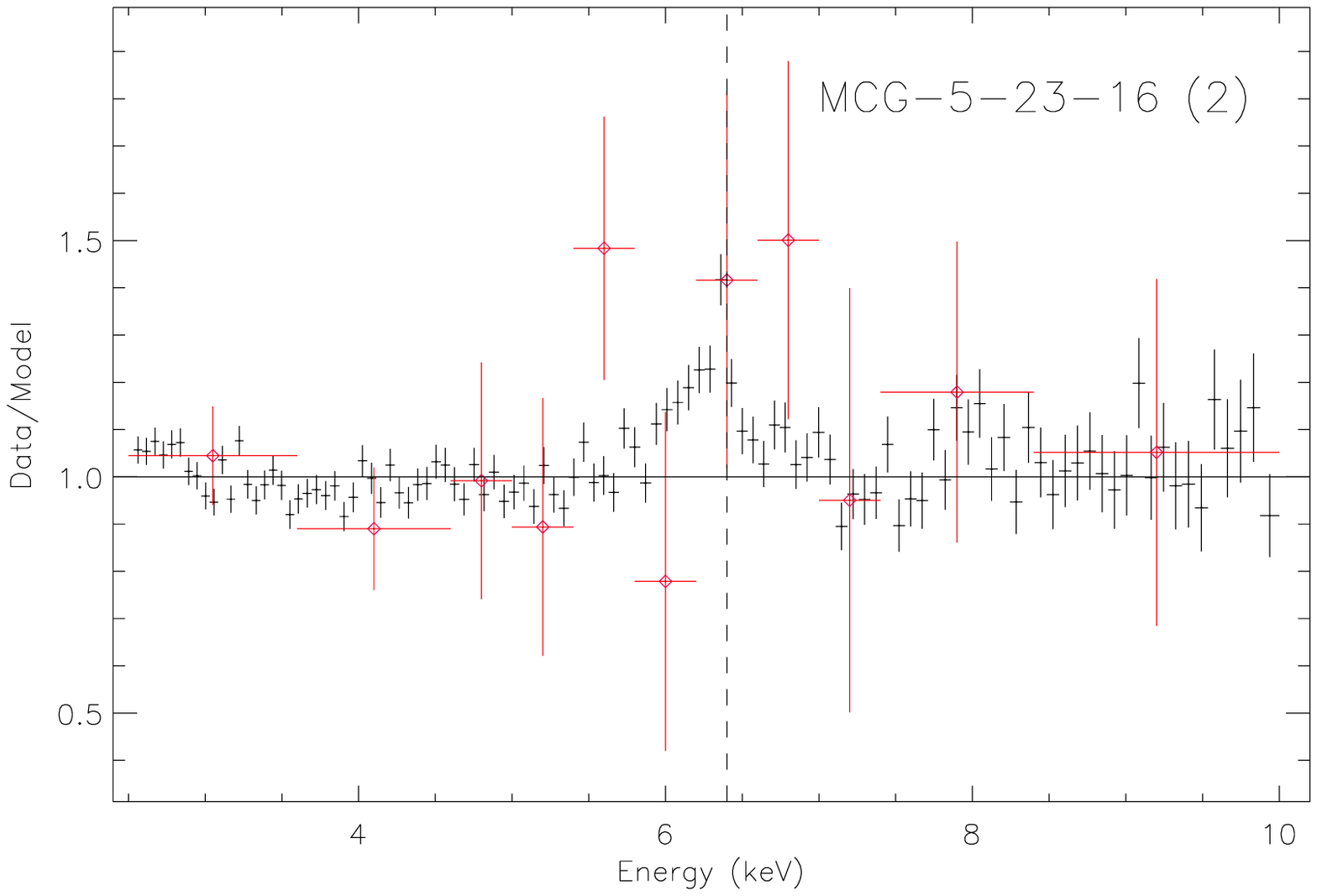}
\includegraphics[angle=0,width=58mm]{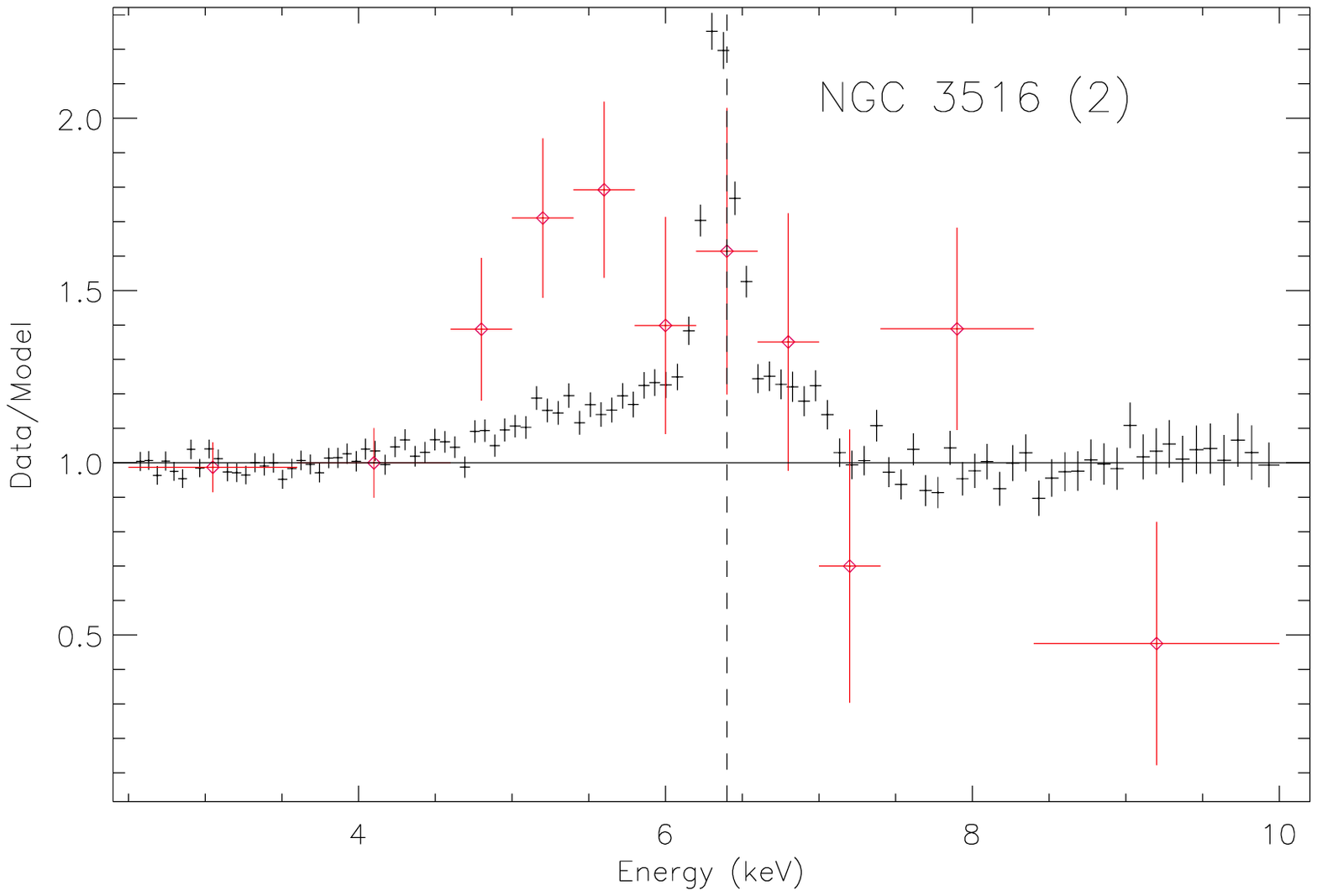}
\includegraphics[angle=0,width=58mm]{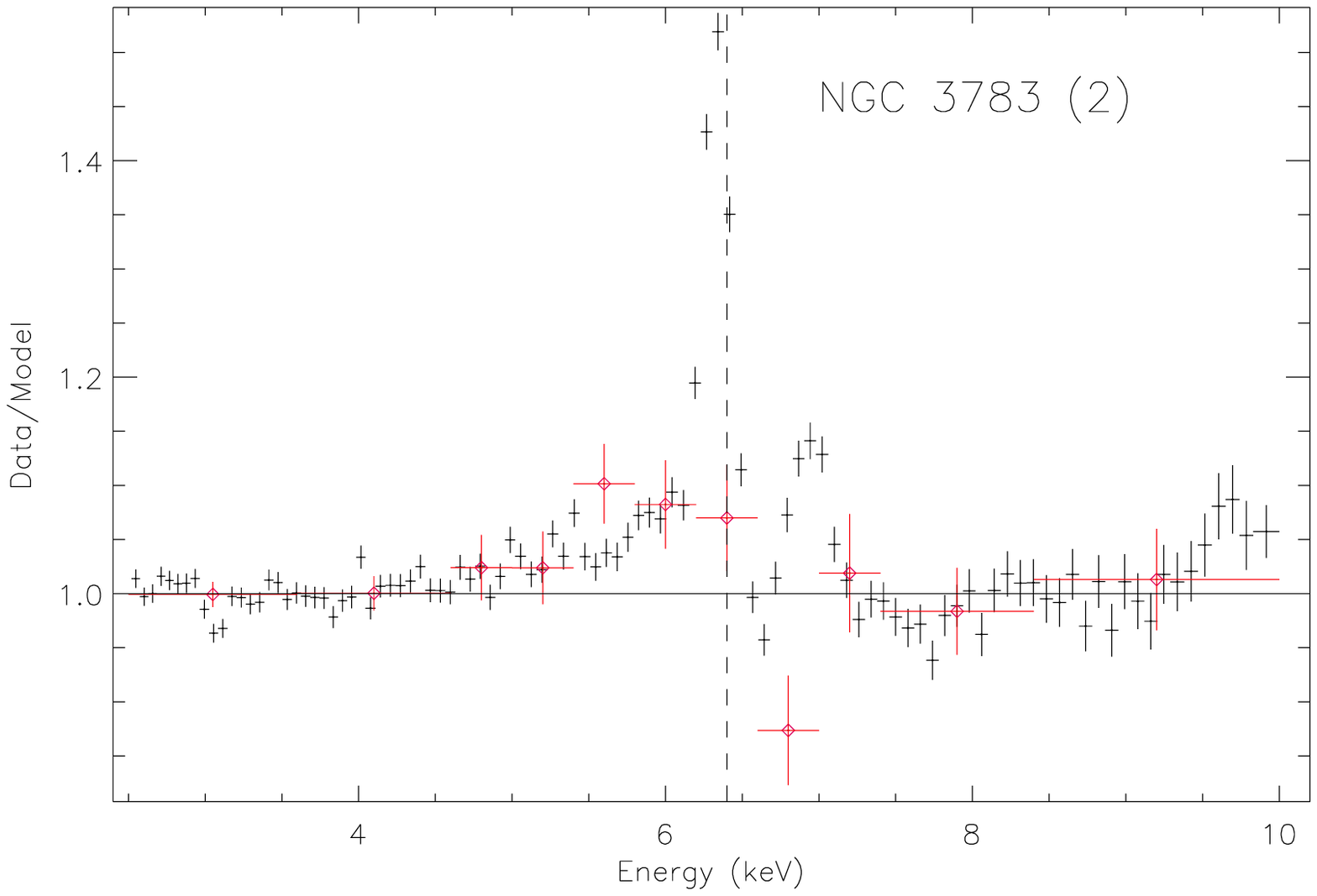}
\includegraphics[angle=0,width=58mm]{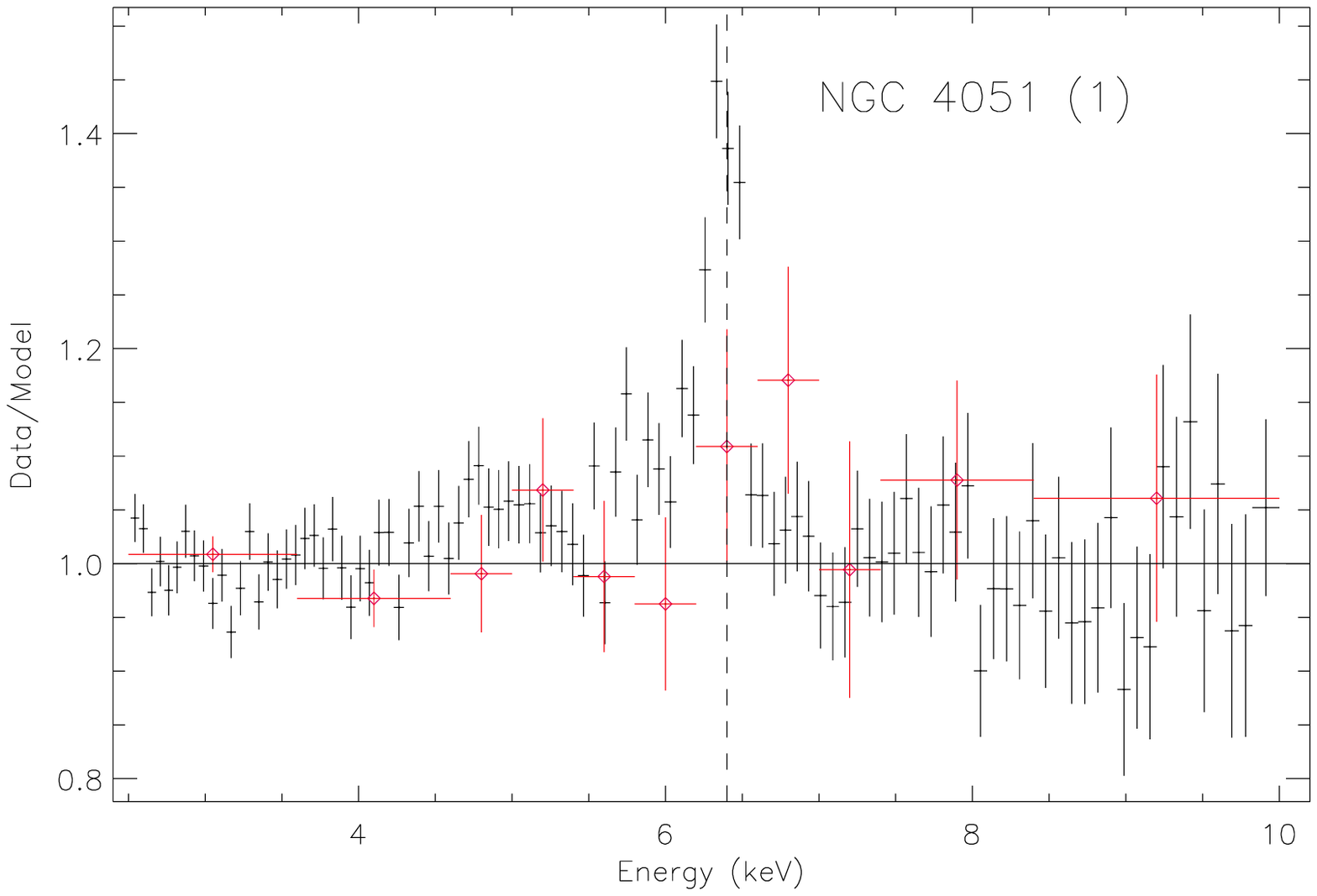}
\includegraphics[angle=0,width=58mm]{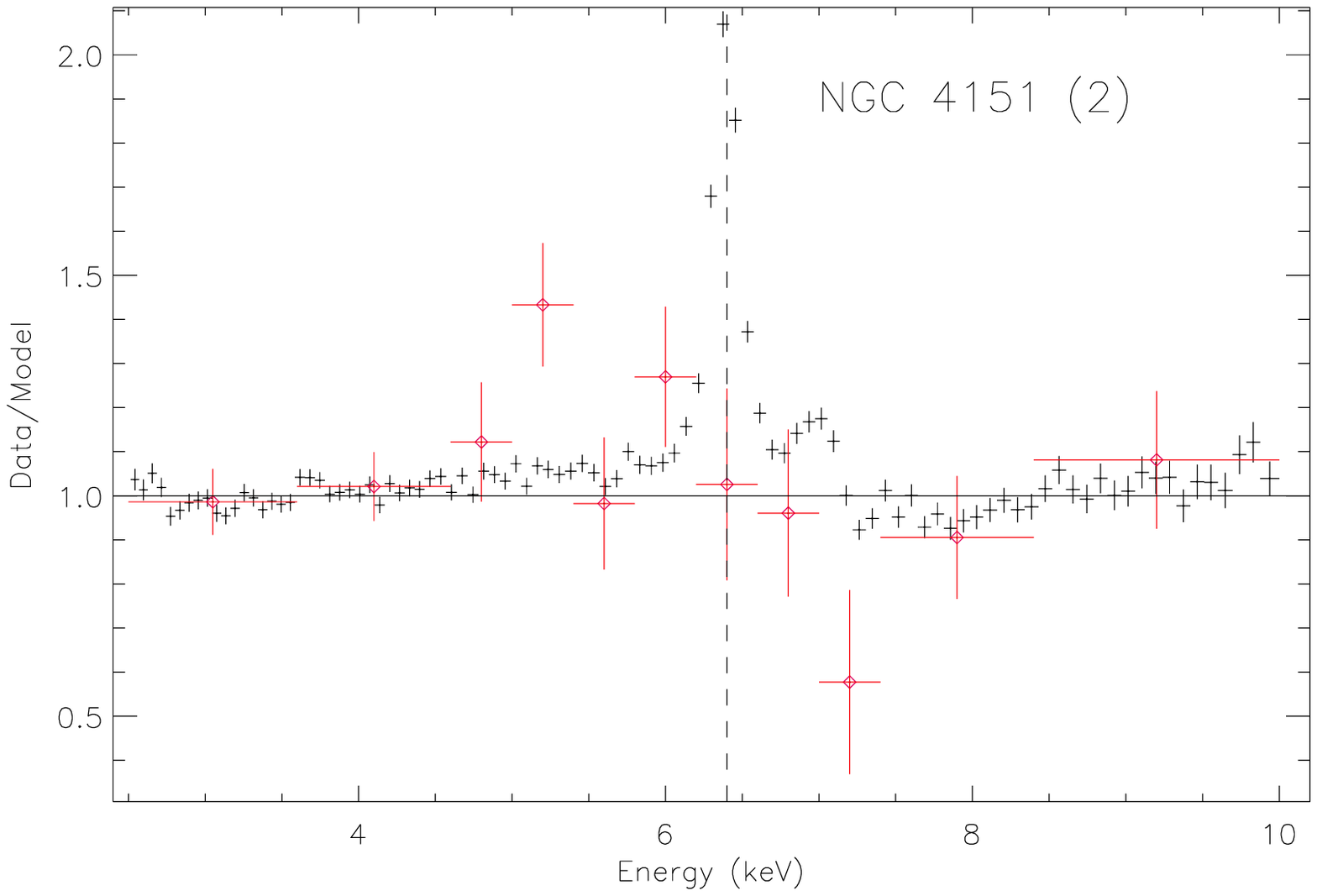}
\includegraphics[angle=0,width=58mm]{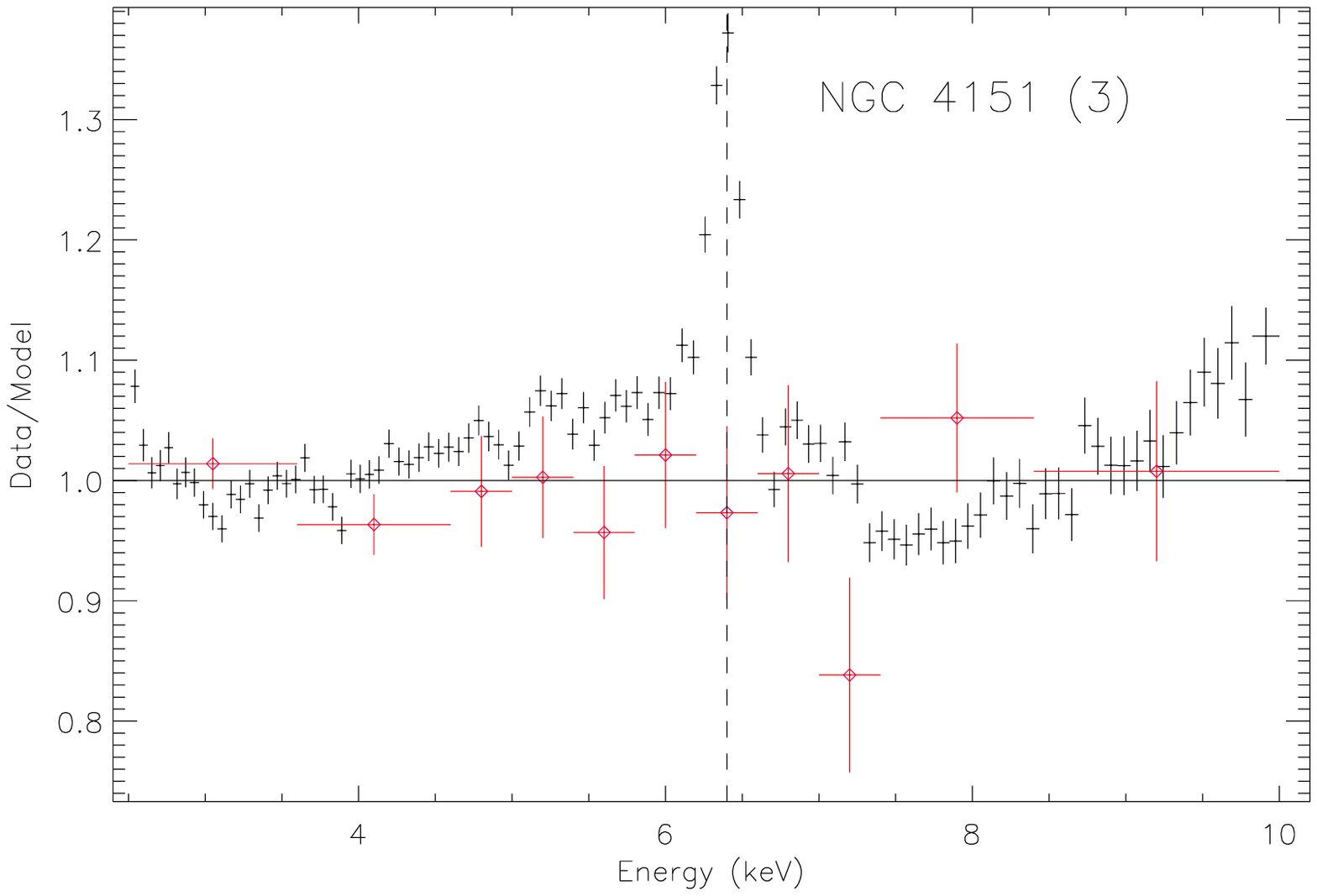}
\includegraphics[angle=0,width=58mm]{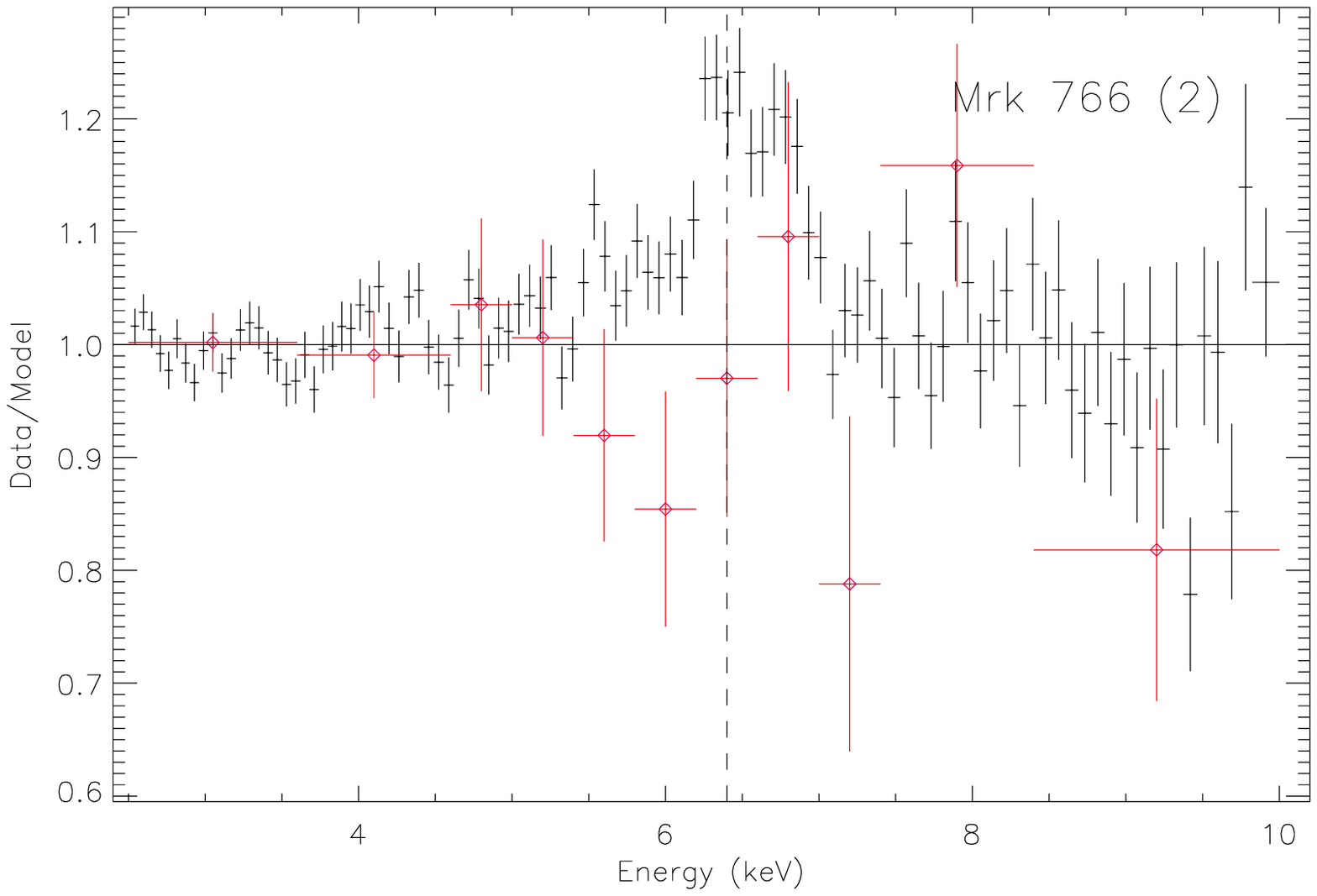}
\includegraphics[angle=0,width=58mm]{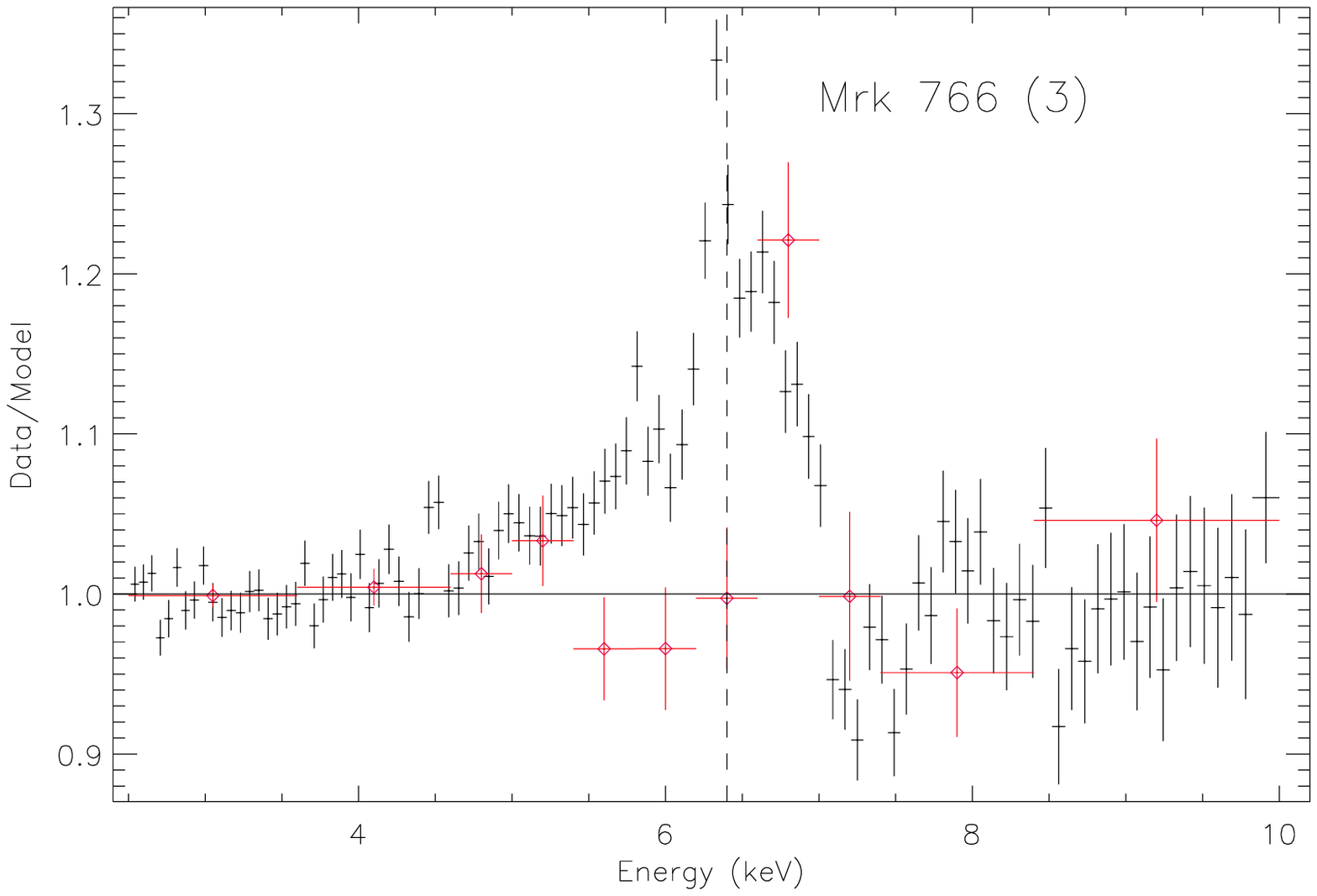}
\includegraphics[angle=0,width=58mm]{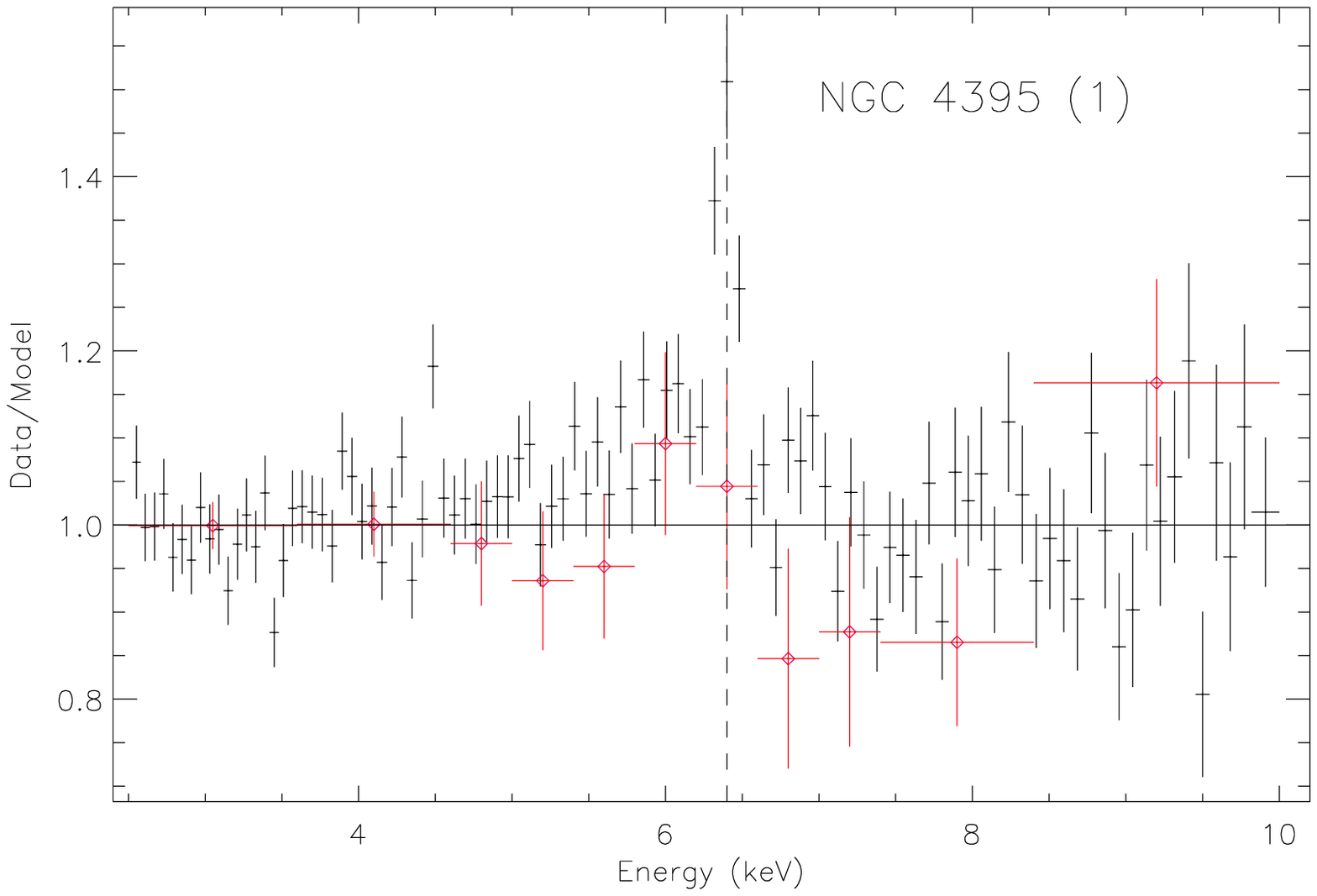}
\includegraphics[angle=0,width=58mm]{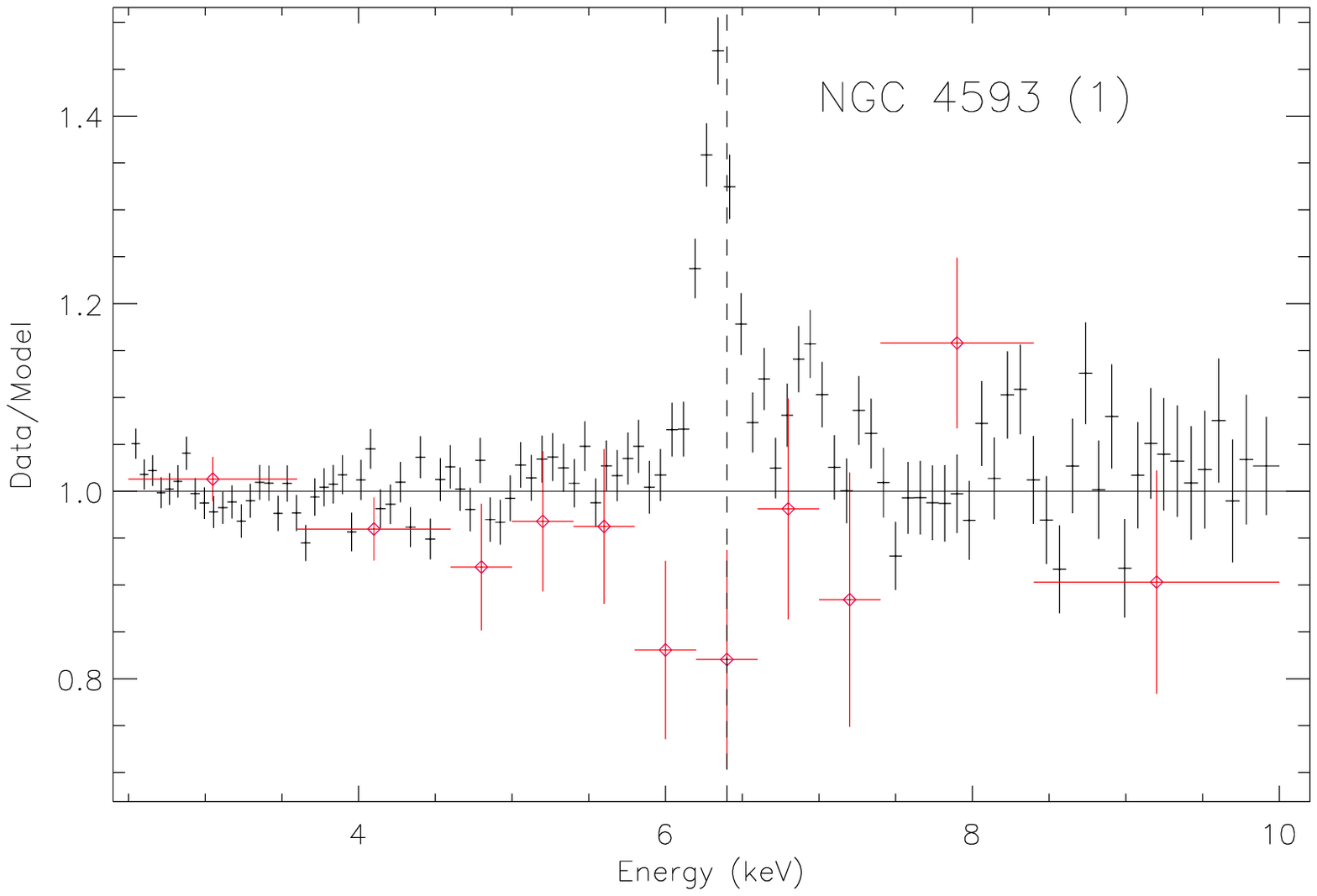}
\includegraphics[angle=0,width=58mm]{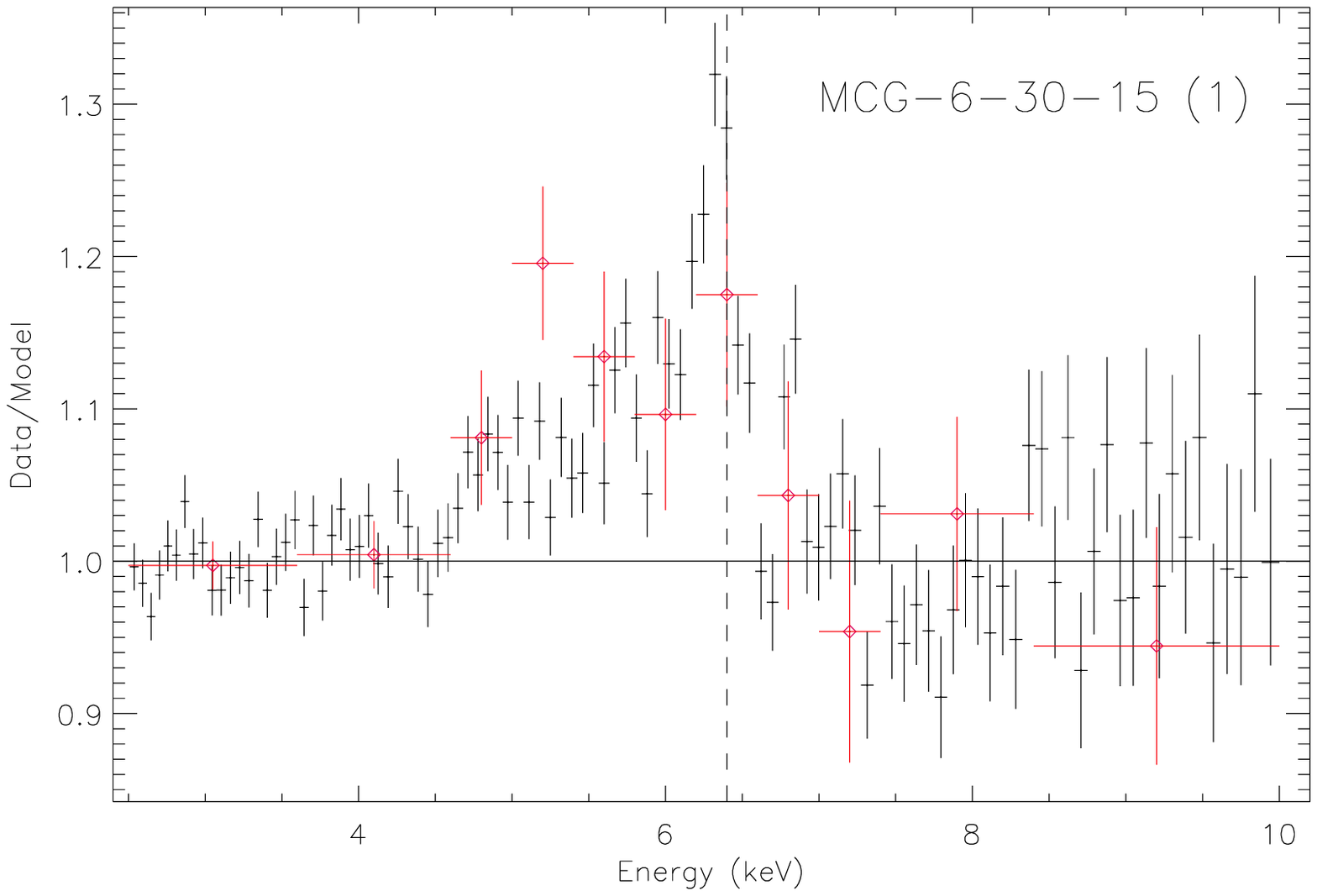}
\includegraphics[angle=0,width=58mm]{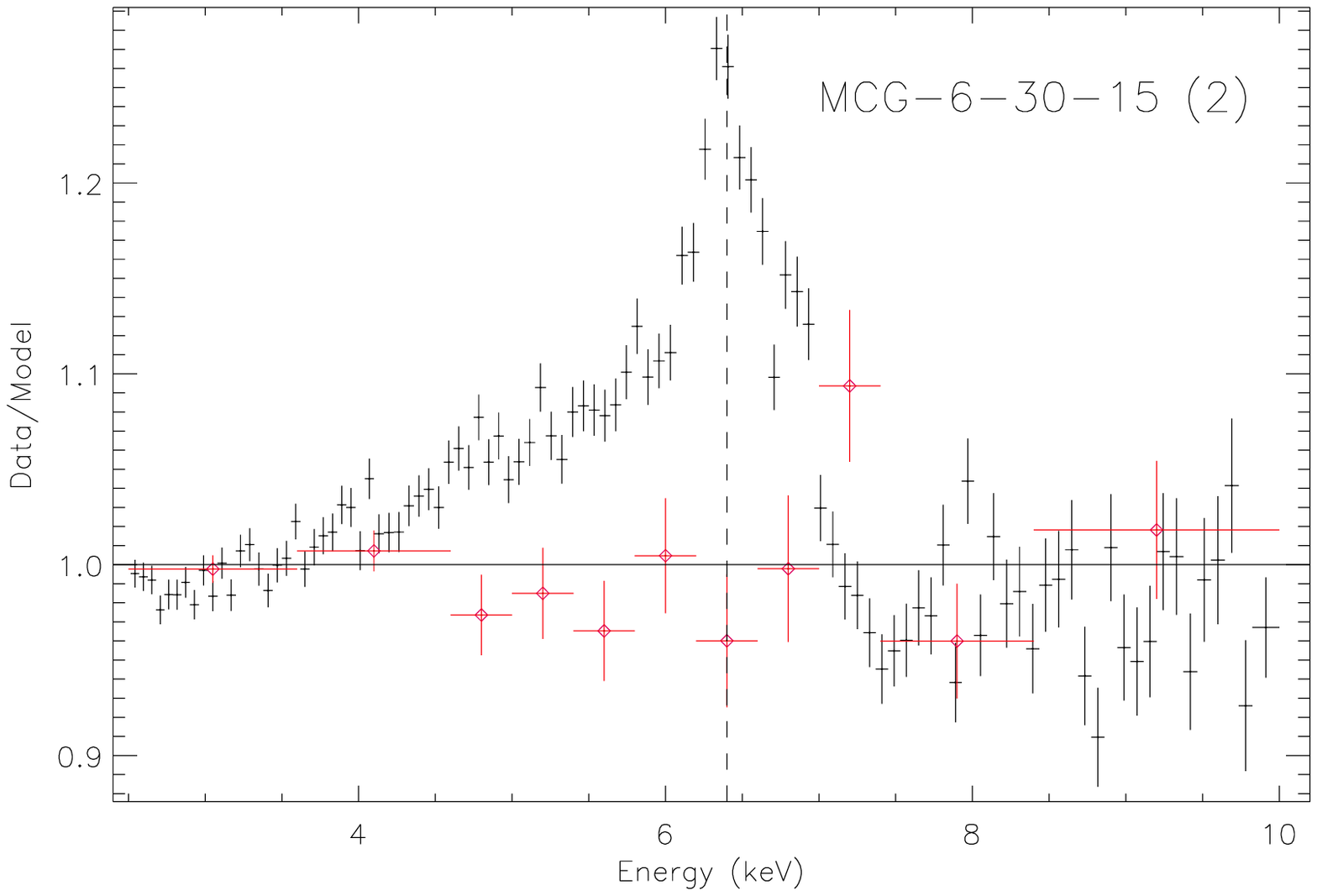}
\includegraphics[angle=0,width=58mm]{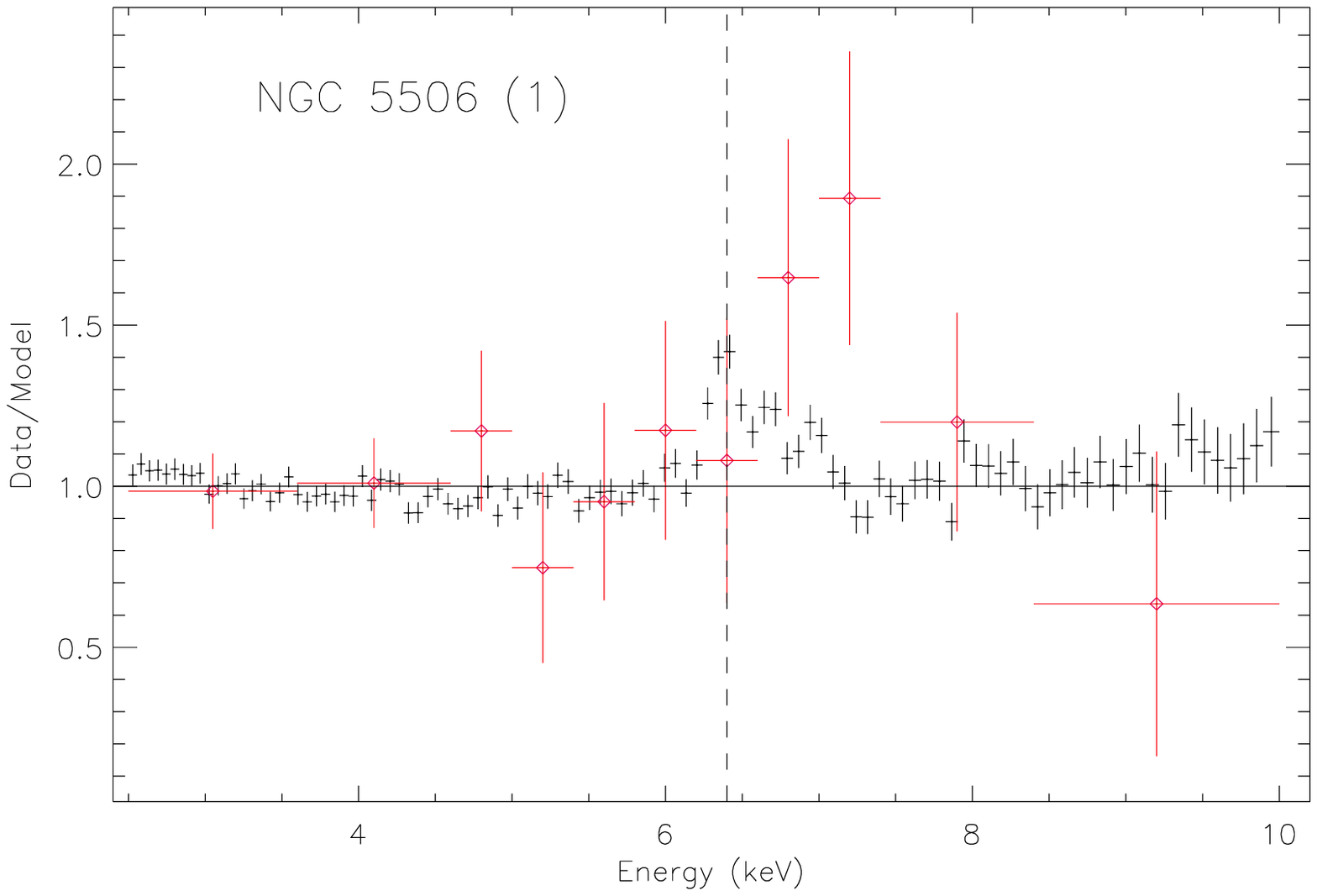}
\includegraphics[angle=0,width=58mm]{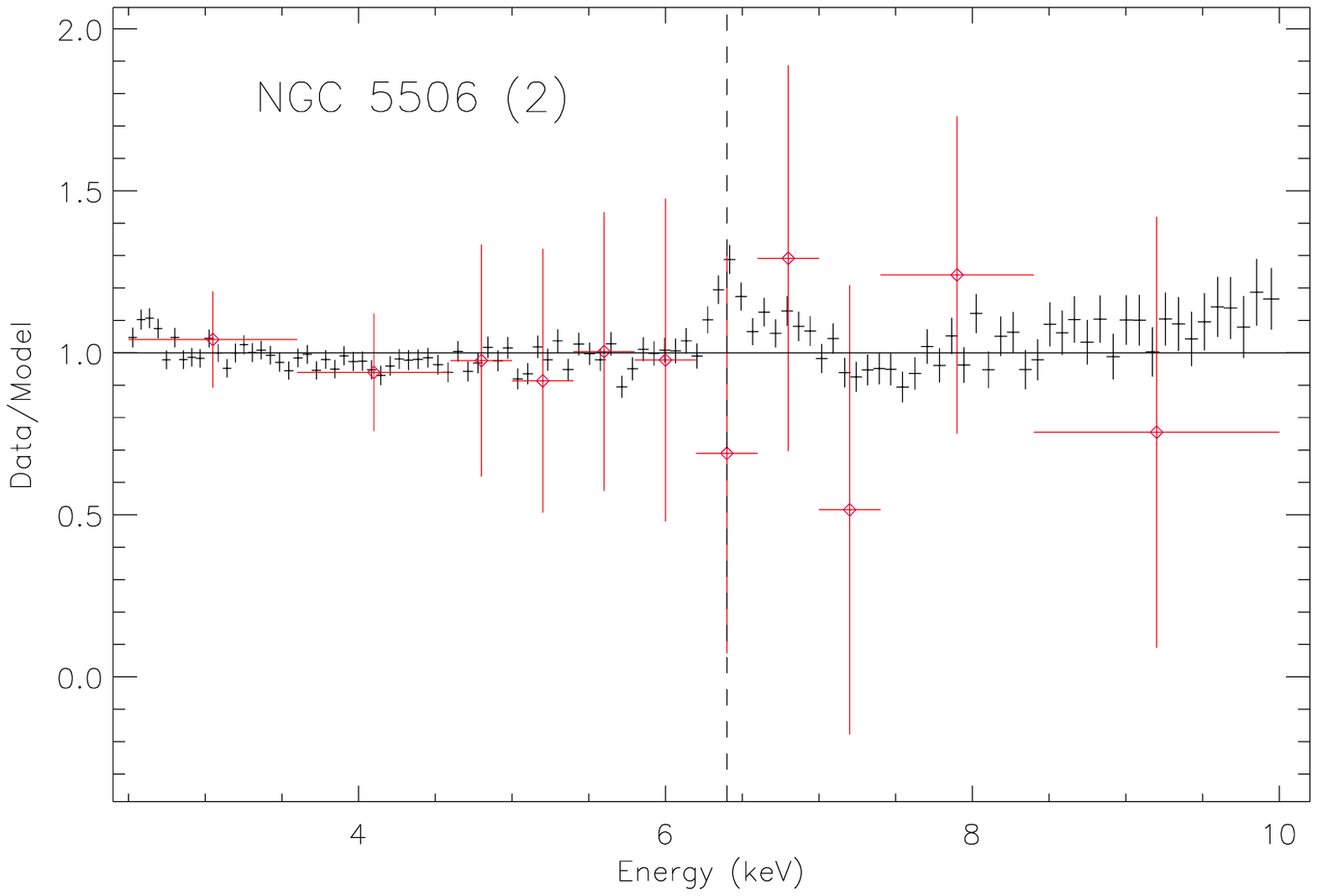}
\includegraphics[angle=0,width=58mm]{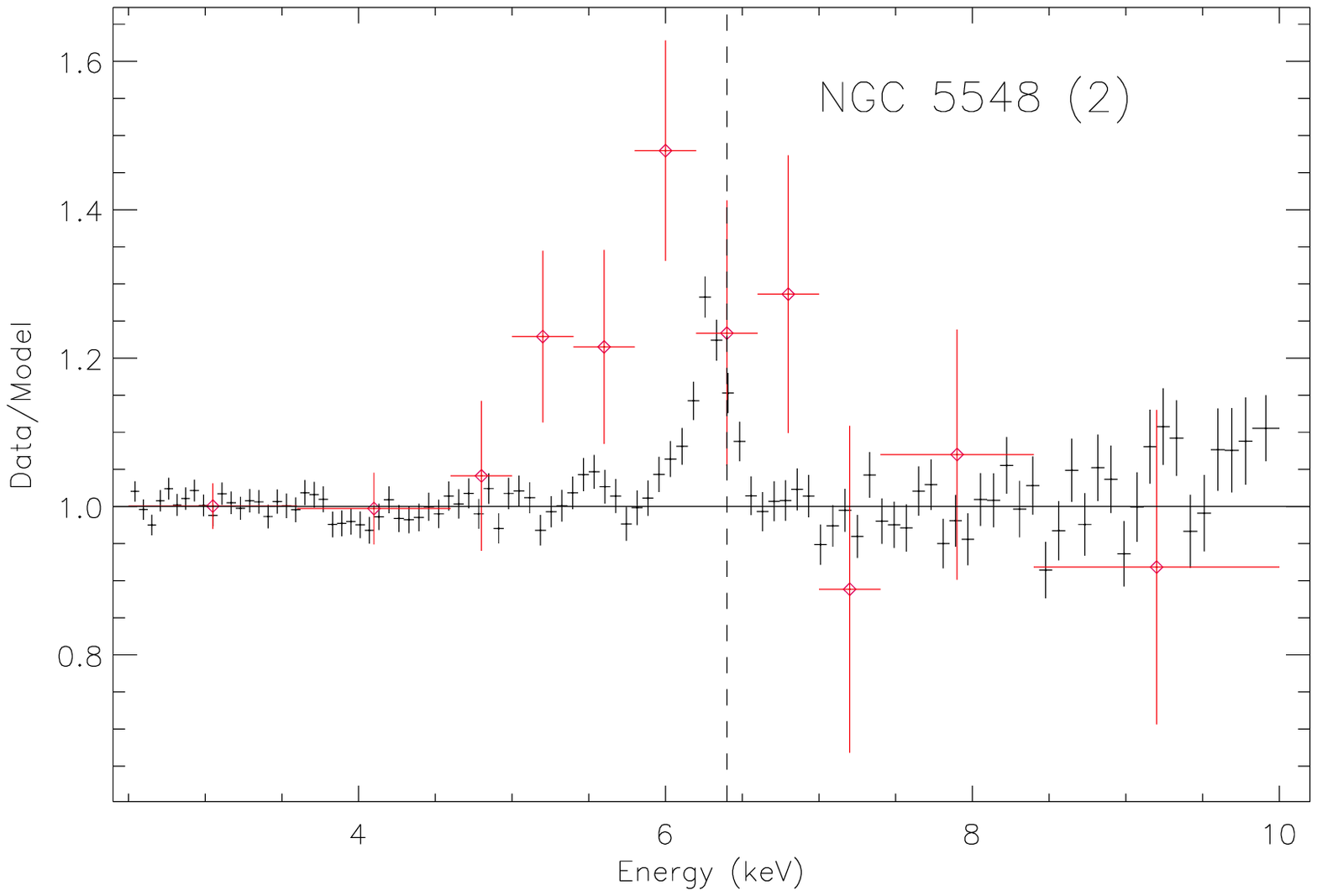}

\caption{Data/model ratios for 18 observations with fully-defined rms spectra. The model is a power-law model with local absorption that has been fitted to the unnormalised rms by excluding the 4.5-7.5 keV interval (thick black error bars, red in the online version). A warm absorber was also included if found necessary by N07 with parameters fixed to their values. The same model has also been fit to the time-averaged spectrum with parameters allowed to vary from the rms fit (thin black error bars). The vertical dashed lines mark the rest-frame energy of 6.4 keV, which is expected for fluorescence from neutral iron.}

\label{fig:u3}
\end{figure*}

\begin{figure*}

\includegraphics[angle=0,width=58mm]{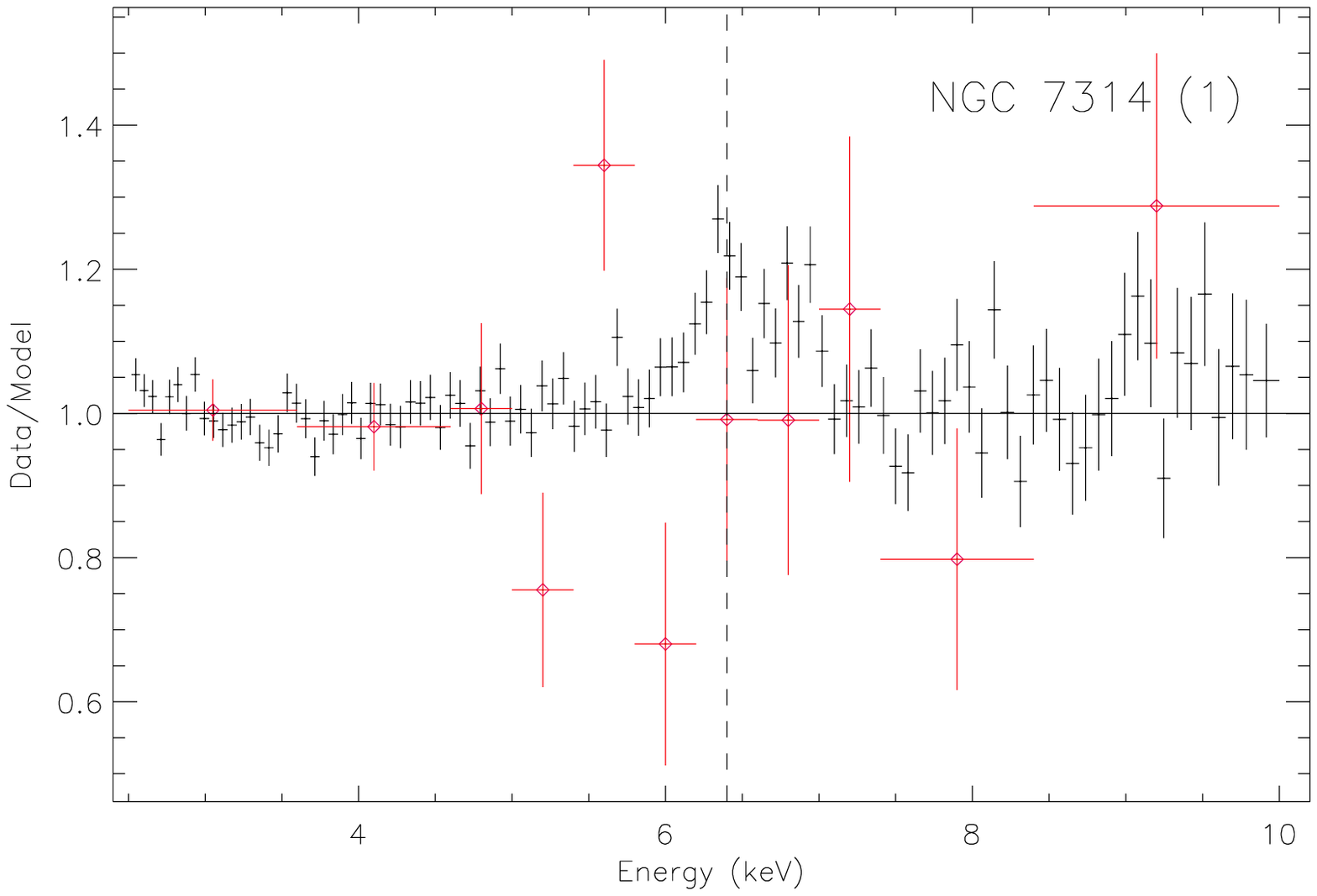}
\includegraphics[angle=0,width=58mm]{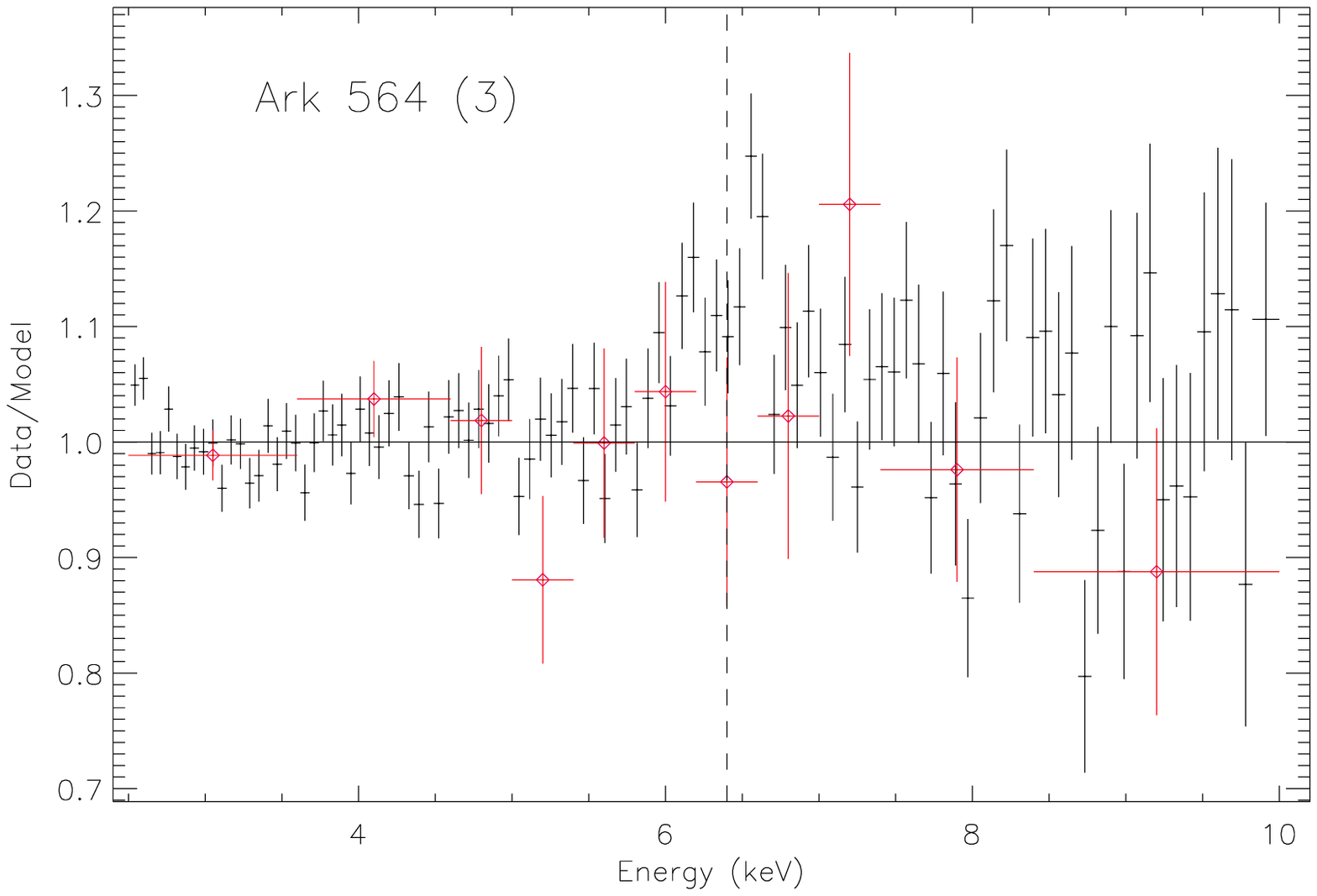}
\includegraphics[angle=0,width=58mm]{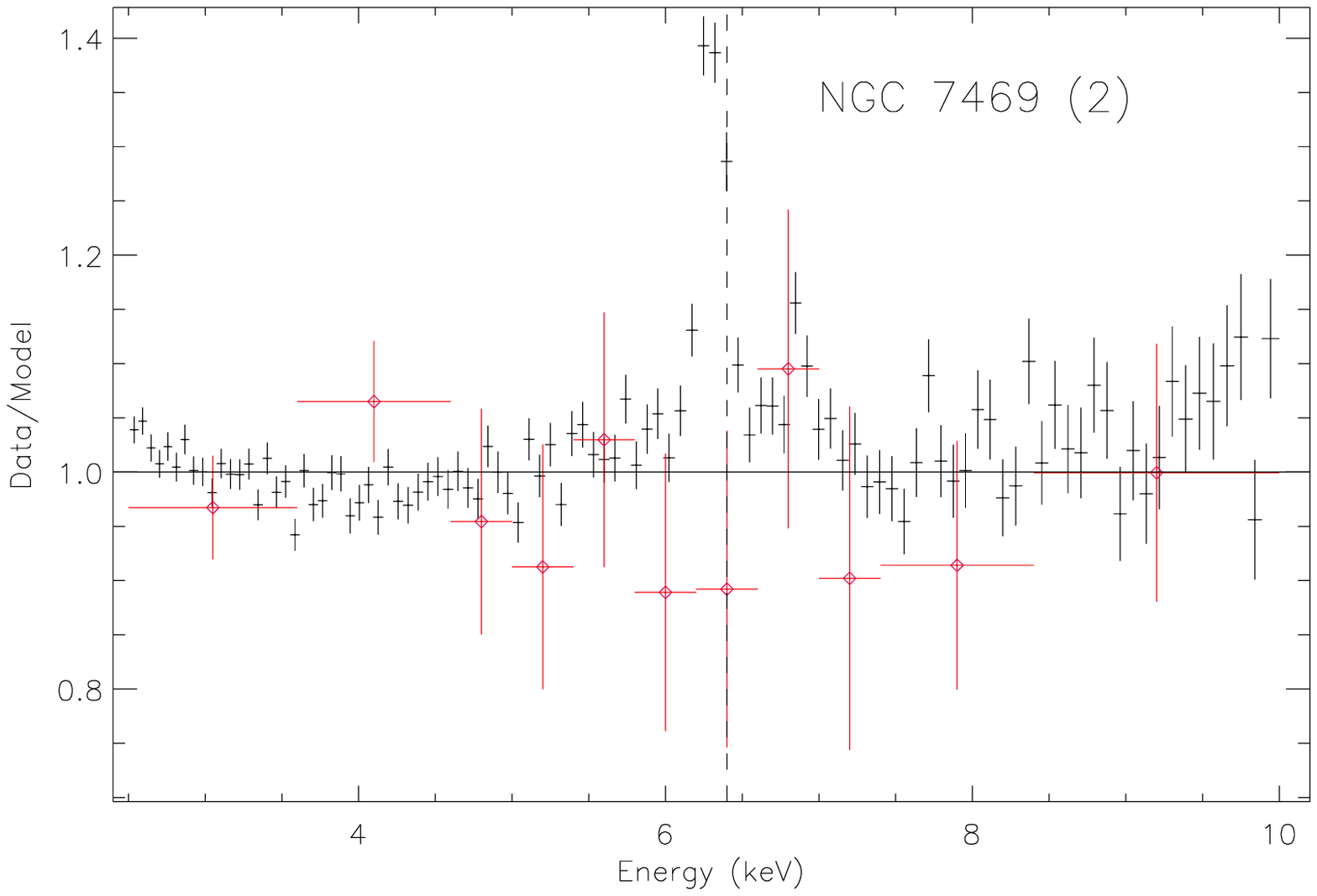}

\contcaption{}

\label{fig:u3}
\end{figure*}

\begin{table*}
\caption{Variability significances of observations with excesses in any of the 7 iron line bands between 4.6 - 7.4 keV. Only excesses with a confidence level of greater than 80\% are included and those greater than 95\% are denoted in bold. }
\begin{threeparttable}

\label{tab:varexcess}
\begin{tabular}{lccccccc}
\hline

& 4.6 - 5.0 & 5.0 - 5.4 & 5.4 - 5.8 & 5.8 - 6.2 & 6.2 - 6.6 & 6.6 - 7.0 & 7.0 - 7.4 \\

& (\%) & (\%) & (\%) & (\%) & (\%) & (\%) & (\%) \\

\hline

MCG -5-23-16 (2) & & & 93.5 & & 83.5 & 84.2 &   \\
NGC 3516 (2) & 87.5 & \textbf{98.5} & \textbf{98.4} & & 89.4 & & \\ 
NGC 3783 & & & \textbf{96.5} & 91.7 & 88.5 & & \\
NGC 4051 & & & & & 80.5 & 87.4 & \\
NGC 4151 (2) & & \textbf{99.3} & & 88.2 & & & \\
NGC 4151 (3) & & & & & & & \\
Mkn 766 (2) & & & & & & &   \\
Mkn 766 (3) & & & & & & \textbf{100.0} &   \\
NGC 4395 & & & & 80.1 & & & \\
NGC 4593 & & & & & & & \\
MCG -6-30-15 (1) & 85.9 & \textbf{99.7} & 94.0 & 85.4 & \textbf{95.2} & &  \\
MCG -6-30-15 (2) & & & & & & & \textbf{98.2}  \\
NGC 5506 (1) & & & & & & \textbf{95.8} & \textbf{100.0}  \\
NGC 5506 (2) & & & & & & &   \\
NGC 5548 (2) & & \textbf{95.1} & 90.1 & \textbf{99.7} & 90.2 & 91.1 &   \\
NGC 7314 (1) & & & \textbf{97.8} & & & &   \\
Ark 564 & & & & & & & 91.0\\
NGC 7469 (2) & & & & & & &   \\

\hline

\end{tabular}

\end{threeparttable}

\end{table*}

\section{Results}

Simple disk models for Fe K$\alpha$ line emission predict that the line should respond rapidly to changes in the illuminating continuum flux so the rms spectrum for such an object would have a power-law and iron line excess that matches the time-averaged spectrum. In contrast, the rms spectra in Fig. 1 display a range of iron line variability profiles and there is no clear overall trend between the time-averaged and rms spectra. 

Half of the sample (9/18 observations) show at least one bin with significant line variability at 95\%. A further four show lower-confidence (80-95\%) excesses only, but in the cases of MCG-5-23-16(2) and NGC 4051 these might well be real, as they exhibit more than one bin with such an excess. Ark 564 contains a single excess bin at 91\% confidence, which is suggestive, but marginal. NGC 4395 contains a single bin at 80\% confidence which we do not consider significant. Of the observations with significant line variability, two (NGC 3516(2) and NGC 5506(1)) show definitive broad line variability, in that there are two adjacent bins with excess variations at $>95$\% confidence. However, a further three observations, NGC 3783, MCG-6-30-15(1) and NGC 5548(2) almost certainly have broad line variability as well, based on moderately significant excesses adjacent to the more significant bins. NGC 4151(2), Mrk 766(3), MCG-6-30-15(2) and NGC 7314(1) show significant, but narrow excesses. 

The diversity of behaviour increases further when the strength of the variability excess is compared to the time-averaged line profile.  The correspondence is generally poor. For example, NGC 3516(2) and NGC 5548(2) have line variability, especially in the red wing, that is stronger than the time-averaged spectrum. MCG-6-30-15(1) has a variable, broad iron line that is as strong as the time-averaged line profile, as does NGC 3783 (although its iron line red wing is not as strong as in MCG-6-30-15). Sometimes, a broad iron line and red wing is seen in the time-averaged spectrum, but the variability excess is narrow, and tends to the blue side of the line. Examples are Mrk 766(3), MCG-6-30-15(2) and NGC 7314(1). NGC 4151(2) is unique in displaying a significantly variable narrow component to the red side of the line, greatly in excess of the time averaged profile. Many observations show no evidence for line variability at all, despite a clear emission (narrow and/or broad) in their time averaged spectra. 


The rms spectra can crudely be classified into three non-exclusive groups. The first class are those with red wing variability, the second, those with blue excess variations, and lastly there are those with little or no evidence for broad or energy shifted iron line variations at all. Here we discuss each class in turn, but before doing so, we consider the 6.4 keV line core, which may well have a different origin to the broad line, outside the accretion disk. 


\subsection{6.4 keV Iron Line Core}

N07 found that a narrow component of the iron line at 6.4 keV was ubiquitous throughout their sample. The fact that it is narrow and found at the non-redshifted energy of fluorescence for neutral iron suggests that it originates at a large distance from the black hole. The most likely location is the molecular torus (Ghiselini, Haardt \& Matt 1994; Krolik, Madau \& Zycki 1994) and such a line would be expected to be constant because any variations in the power-law continuum will be averaged out over the light-crossing time, which is expected to be several years for a typical pc-scale torus. This agrees with the fact that no observation other than MCG-6-30-15(1), which has a broad variable iron line excess, shows a significantly variable excess in the 6.2-6.6 keV bin. In all cases, the variability at 6.4 keV is suppressed compared to the time average spectrum. Clearly the narrow core of the iron line is far less variable than the continuum, supporting an origin in distant material. 

\subsection{Red Wing Variability}

A number of observations in Table 1 show variability excesses that extend below 6.4 keV. In many the excesses are observed in multiple, consecutive bins. In some cases, the red wing is as variable as the continuum, e.g. NGC 3783 and MCG-6-30-15(1), but in others, the line is more variable: e.g. NGC 3516(2) and NGC 5548(2). The observation that the red wing is as variable as the continuum is consistent with the simplest interpretation of disk reflection. 

The excess in the rms spectrum of NGC 3783 is similar in strength to the broad iron line in the time-averaged spectrum, so in the accretion disk model, the line and continuum vary together. N07 found an ionized H-like line at 6.97 keV that is not seen to be variable in the rms spectrum. This offers support for the speculation of N07 that this originates from hot gas filling the torus, which is not expected to be variable. Turning to MCG-6-30-15, N07 found a strong requirement for a blurred accretion disk component in the time averaged spectrum (as in several previous studies, e.g. Tanaka et al. 1995; Fabian et al. 2002). The variability line profile matches well with the time-averaged line profile in the first of our observations, MCG-6-30-15(1), which follows the simple disk interpretation. However, there is a large change between this and the variability profile for the second observation (see Section 3.5). NGC 3783 and MCG-6-30-15(1) are the only observations in our sample which both show clear evidence for broad line variations, and an amplitude consistent with the continuum, although there are others where the signal-to-noise ratio prevents definitive conclusions.



There are four observations which show variability of the iron line redward of the core that exceeds that of the time averaged spectrum. NGC 3516(2) and  NGC 5548(2) are the clearest cases, along with NGC 4151(2) and MCG-5-23-16(2) showing narrower or more marginal excesses. If the red wing is indeed more strongly variable than the continuum, it is a compelling indicator that relativistic effects are in play (Ponti et al. 2004; Miniutti \& Fabian 2004). Curiously, however, N07 did not find evidence in either NGC 5548(2) or NGC 4151(2) for strong gravitational effects in the time averaged spectra. NGC 5548(2) is the most puzzling case, with a non-relativistic iron line but a strong, broad variable excess that extends down to 5.0 keV, a much lower energy than the time-averaged profile.  MCG-5-23-16(2) and NGC 3516(2) are different in that they possess relativistically broad iron lines, as well as having strong variability excesses across the red wing. This means that the iron line is more variable than the continuum and is likely to be explained by relativistic effects such as beaming or emission from iron moving with a large bulk velocity. While most of the excess variability is always seen in the red wings, in MCG-5-23-16(2) and NGC 5548(2) the variability also extends up to 7.0 keV.

\subsection{Blue Excesses}
\label{sec:blue}

The second class of observations exhibit the majority of their variability blueward of the line core, and include Mrk 766(3), MCG-6-30-15(2), NGC 5506(1) and the more marginal cases of NGC 4051 and Ark 564.


In Mrk 766(3), the variability excess in the 6.6-7.0 keV bin is narrow compared to the broad iron line and at a higher energy than the narrow core at 6.4 keV. The time-averaged spectrum is fit well with only an accretion disk model, $\chi^{2}$ =  94.8 for 93 degrees of freedom but the rms spectrum suggests that the iron K$\alpha$ region can be split into two bands of high and low ionisation. The high ionisation band varies with the continuum, while the low ionisation band representing disk emission is nearly constant. NGC 5506(1) also displays blue iron line variability which extends up to 7.4 keV. N07 found that the time-averaged spectrum could be explained by a blend of high ionization lines or disk reflection. The emission lines are at 6.4 and 6.7 keV, while Bianchi et al. (2003) have also found evidence for an H-like line at 6.96 keV. It is unclear if the variability excess relates to the high ionization lines lines as there may also be variability above 7 keV. In contrast to Mrk 766(3), the relative lack of variability below 6.4 keV is due to no broad wing being present rather than a lack of variable red wing emission.

MCG-6-30-15(2) largely has a flat unnormalised rms spectrum apart from a significant excess in the 7.0-7.4 keV bin. No emission or absorption is found in this energy band by N07, but Fabian et al. (2002) do find complexity around 7.0 keV that can be explained by an edge at 7.38 keV due to thick material partially covering the line of sight. The variability could then be due to changes in the covering rather than being attributable to the accretion disk. Alternatively, the complexity could be due to H-like recombination emission line at 6.97 keV, but then it is puzzling that there is no variability in the 6.6-7.0 keV bin. Ark 564 was found to be equally well fit by a blend of narrow lines (He- and H- like iron lines) and a multizone warm absorber as by a blurred reflector. The only variability excess is in the 7.0-7.4 keV bin so it does not appear to be associated with either of the narrow lines fitted to the time-averaged spectrum by N07. This may argue in favour of the blurred reflection interpretation. NGC 4051 is fit well by an accretion disk model but Ponti et al. (2006) have found that the iron line is not variable in the same observation of NGC 4051 when it was in a high flux state. This agrees with the fact that the rms spectrum shows no variability below 6.2 keV and only marginal variability above. Interestingly Ponti et al. find that another observation at lower flux did possess an iron line that varied with the continuum. Pounds et al. (2004) have also found an absorption line at 7.1 keV that is variable. However, the marginal excess in the rms spectrum  at a lower energy than this, so the origin of this variability is unclear.


\subsection{Constant iron lines}
\label{sec:constant}



A large group of observations show clear evidence for a red wing of the iron line in their time averaged spectra, but have no evidence for variability of that emission component in the rms spectrum. This group can be split into NGC 5506(2), NGC 4151(3) and Mrk 766(2), which have no evidence for variability across the entire iron line region and MCG-6-30-15(2), Mrk 766(3) and NGC 4051, which have blue but no red variability. NGC 7469(2) and NGC 4593 are not variable across the iron line region, but neither was found to possess an obvious broad iron line component. 

The apparent lack of red wing variability in NGC 5506(2) and Mrk 766(2) could be due simply to poor statistics. Arguably the most interesting cases are where there is a clear broad red wing of emission, but a flat rms spectrum. The most obvious examples are MCG-6-30-15(2), NGC 4151(3) and Mrk 766(3). These observations are those where the reflection component appears constant while the power-law continuum is the only variable component.

Three observations, NGC 4593, NGC 7314(1) and NGC 7469(2) appear to have dips in the iron line region of their rms spectra. Dips do not mean anything physically and may be indicative of incorrect modelling of the continuum. The time-averaged spectrum of NGC 7469(2) was not fit well with a reflection component, though there is a significant improvement when a blurred reflector is included. N07 proposed that this could be due to a black hole with an extreme spin that blurs the line to such a large extent that it can be difficult to separate from the continuum. This could make the continuum outside the iron line region concave, and thus the use of only a power-law in fitting the continuum is too simplistic. Similarly, NGC 4593 was also found to have a hard tail so a better model for the continuum is needed. Unfortunately, not enough bins are available to do this for these two observations. NGC 7314(1) has both variability dips and an excess across the red wing while the rest of the iron line does not appear variable. The time-averaged spectrum is well fit with a disk reflection model but clearly the rms spectrum is different. This observation was not found to have any obvious additional continuum complexity, which makes the extremes in variability more difficult to understand.

\subsection{Objects with multiple observations} 

There are four objects where we are able to compute rms spectra for more than one observation: MCG-6-30-15, Mrk 766, NGC 5506 and NGC 4151. They provide an opportunity to test if the variability of the iron line has changed with the time-averaged spectra or has remained the same. Arguably one of the most puzzling results of our study is that the rms variability properties appear to change in all of these cases.

The largest difference in variability profile of the iron line across a pair of observations occurs with MCG-6-30-15, as already discussed. The second observation is flat below 7.0 keV, which agrees with other studies of this object (mostly based on this long observation; Fabian et al. 2002; Larsson et al. 2007) where the line has been proposed to be nearly constant while the power-law is the main driver of variability. On the other hand, the earlier observation, MCG-6-30-15(1) shows a highly variable red wing, originally noted by Ponti et al. (2004). This variability behaviour is inconsistent with the two-component model that has been popularised based on this object. Furthermore, the only variable bin in the rms of MCG-6-30-15(2) is the 7.0-7.4 keV bin which is not variable in the first observation. 

The other objects with multiple observations, while less statistically compelling, also appear to show changing variability behaviour. Like MCG-6-30-15(1), NGC 4151(2) shows evidence for excess variability between 5.0-6.2 keV, whereas NGC 4151(3) is flat. Interestingly, N07 found that the third observation, NGC 4151(3) strongly requires a broad and relativistic iron line but the first two observations of this object, while being consistent with a broad line, did not need one. The extension of the variability excess down to below 5 keV in NGC 4151(2) points towards relativistic effects being important, despite clear evidence to the contrary in the time-averaged spectrum. Similarly, the third observation would be expected to show some variability across the iron line region due to the region of line formation being close to the black hole.

NGC 5506(1) shows a clear excess of variability in the blue wing, with NGC 5506(2) not showing any variability. The time-averaged spectra are very similar to each other, so this is very puzzling. There is no red wing variability in either observation of Mrk 766, but there is blue variability in Mrk 766(3) that is not significant in Mrk 766(2). In this instance, this may simply be due to poor statistics in the case of Mrk 766(2). 

\section{Discussion}

We have presented the rms variability spectra of a sample of 18 observations of 14 Seyfert galaxies observed by \xmm, concentrating on their iron K$\alpha$ line variability. Perhaps the most striking result of our study is the extraordinary diversity in the rms spectra. Different objects clearly display different variability properties around the iron line. The rms spectrum is typically also different to the time-averaged spectrum, sometimes showing suppressed variability around the iron line, and enhanced variability in other sources. Finally, objects observed at more than one epoch show changes in their rms spectra, indicating different variability modes at different times, even in the same object. The diversity in the rms variability at the iron line throughout the sample shows that it is unlikely that there is a universal line formation mechanism. Indeed our results are extremely difficult to explain in any single theoretical framework. 

One robust and easily-interpreted result of our study is that, in all observations, we find a lack of variability at the neutral iron line energy of 6.4 keV. Nearly all of the objects were found by N07 to contain a narrow line at this energy. The lack of variability of the line core in the rms spectra supports the interpretation that it arises from material far from the central black hole, for example in the molecular torus. The core is expected to be constant because variations in the power-law continuum will be averaged out over the light crossing time of the torus and therefore emission from the torus should not have any effect on the rms.

The broad iron line emission and its variability are far more puzzling. The simple disk interpretation for the broad line is that variations should follow those in the continuum, as the line emitting region is close to the production site of the power-law continuum, the corona. This would result in an rms spectrum that shows a variable broad line that matches the time-averaged spectrum. This is seen only in a very few objects. Variability in the broad emission is not uncommon in the sample, however, being present in 6/18 observations. This includes at least two observations where variations in the red wing exceed those of the continuum, indicative that in these objects, beaming, light bending or other relativistic effects may be in play.

Some of the objects that display broad iron line variability have been suggested to possess absorption layers that can mimic the red wing e.g. MCG 6-30-15(1) (Miller et al. 2008). A partial covering absorber, when combined with distant reflection, can produce a hump in the spectrum between 3-7 keV, which is then taken to be an emission feature. This requires no relativistic component as the absorption zones are $>100 r_g$ from the black hole and the weak narrow iron line arises completely from the distant torus. Another example is NGC 3783, where there has been some controversy over whether or not a broad red wing is present. Reeves et al. (2004) suggested that much of the broad red wing may be an artefact of a warm absorber adding curvature to the continuum, although they and N07 both concluded that some disk reflection was also present. None the less the spectrum is very complex, with two zones of ionized gas, a strong H-like emission line and a blurred Compton reflection component. The effect of absorption on the rms is not completely clear but if the red wing is a product of partial covering absorption or distant scattered components, then the rms variability of the red wing cannot be larger than that of the continuum. This is compatible with NGC 3783, but disagrees with the suggestion that thick warm absorbers can mimic the red wing in NGC 3516(2) (Turner et al. 2005) as the variability of the iron line is much stronger than the continuum. The rms spectrum is thus better explained with the blurred reflection model (N07) if relativistic effects are enhancing the variability of the red wing. A similar argument can be applied to the observations where complex absorption has been suggested to solely cause spectral curvature in the red wing (e.g. Reeves et al. 2004) but a flat rms spectrum reveals a non-variable broad line and constant disk reflection. For example, Schurch et al. (2003) have suggested that the red wing in NGC 4151(3) might be accounted for purely by complex absorption. In this case the excess photons seen in our time-averaged spectrum between 4-6 keV are continuum photons, which should naturally show the same variability as the remainder of the 4-6 keV continuum, over which the red wing is seen. We are able to reject this possibility using the flat rms spectrum for NGC 4151(3) and conclude that when invariant, the broad red wing must therefore be a distinct additional component, almost certainly from reflection.

%

A flat rms spectrum corresponds to a fractional (i.e. normalised) variance spectrum with a dip and has been found previously in, e.g., MCG-6-30-15 (Vaughan \& Fabian 2004). This has been taken as evidence of a decoupling between the continuum and line. With MCG-6-30-15, the peak of the reflected emission forms close to the black hole, so for the reflection component to not be strongly associated with the direct power-law component, the intrinsic X-ray emission must be anisotropic. Gravitational light bending has been proposed to explain this as the anisotropy can result from the gravitational field of the black hole bending power-law photons away from the observer and towards the disk. This light bending is most severe when the height of the corona over the disk is small. Instead of reaching us, the continuum photons will be reflected off the disk which can enhance the disk component. As the coronal height increases, less photons are deflected onto the disk and so more reach us directly. Miniutti \& Fabian (2004) have calculated how the intensities of the direct and reflection components may vary with the corona height in a simple toy model. They find that at a low coronal height, both vary together which would give an rms spectrum with an excess that matches the time-averaged iron line profile e.g. MCG-6-30-15(1). As the height increases further, another regime is reached where the direct flux varies by a factor of 3 or 4, but the reflection component changes by less than 10\%. Thus this could correspond to flat rms spectra such as that of MCG-6-30-15(2). As the two observations of MCG-6-30-15 are separated by a year, the light bending interpretation suggests that there is a large change in the coronal height within a relatively small amount of time. The observations where the variability excess is stronger than the red wing of the time-averaged iron line could also be explained by gravitational light bending if relativistic effects such as relativistic beaming enhance the variability of the disk reflection component. However, NGC 4151(2) and NGC 5548(2) have the strong excesses but do not have iron lines that are relativistically broadened as would be expected if this interpretation is correct. Along with MCG-6-30-15, the other three objects with multiple observations, NGC 4151, Mrk 766 and NGC 5506 show a change in the variability at the iron line between their two observations. This demonstrates the complexity of variability in Seyfert 1s as there is not only a range of behaviour found between objects, but also within objects themselves. The implication is that the inner accretion flow is highly unstable and chaotic, with changes in illumination and/or geometry occurring on relatively short timescales (months-years). Relativistic transient lines could be a cause for the difference in variability of the iron line region between multiple observations especially if they are strongly variable. Mrk 766(2) was found by Turner et al. (2004) to have such a spectral component between 5.6 - 5.75 keV but this bin shows no excess rms variability. NGC 3516(2) was also found to have narrow transient iron lines at 5.6 keV and 6.2 keV (Turner et al. 2002) and these bins are are variable in the rms spectra though it is likely that this is mainly due to the variable broad iron line. Furthermore, the lines are narrower than the 0.4 keV energy bins of the rms spectra so the variability of the transient line might be washed out. Longer observations with high signal-to-noise ratio are needed to investigate the variability of transient iron lines.

There are alternatives to gravitational light bending in explaining a flat rms spectrum. The reduced variability of the line could also be due to the inner disk being absent or suppressed so the line is unable to form close to the black hole where it would be more variable. This could be because the inner disk is highly ionized, radiatively inefficient or truncated. The bulk of the observed line flux would then arise from the outer region of the disk. However, many of the observations with flat rms spectra have very broad iron lines, with a profile indicating that the emission is concentrated close to the black hole, so non-variability of the line should not be due to the inner disk emission being missing.

The origin of the narrower variability excesses, typically seen blueward of the line core, are also difficult to explain. When they are seen there is generally no variability at the red wing, so if the blue emission is the high energy wing of the diskline, it is hard to understand why the red part is not also variable: the red wing of the line should originate closer to the black hole where strong gravitational redshifting effects are present. An alternative is that the blue variability is due to absorption or emission by ionized iron close to the central source, but not too close so as to be broadened. This has been suggested for Mrk 766(3) by Miller et al (2006). They split the iron K band into high and low ionisation bands and find only the former is variable. This agrees with the rms spectrum showing an excess in the 6.6-7.0 keV bin that is as large as the time-averaged component. Mrk 766(2) does not show such a strong excess at a similar energy and Pounds et al (2003) do not find any complexity around 7keV in the spectrum. In addition there are also observations with clear evidence for relatively narrow, ionized lines in their time-averaged spectra (e.g. NGC 3783, NGC 4593) which show no evidence for variability, indicating a more distant origin for these lines. Perhaps most puzzling of all are cases like NGC 5506(1) (or NGC 5548(2)) which show strong blue (or red) wing variability without any obvious associated features in the time averaged spectrum. Even if absorption or emission by photoionised gases does contribute, simple models would find it difficult to explain the varying strengths of the excesses. NGC 4151(2) is unusual in that it shows a narrow excess redward of the line core. The evidence for a red wing the time-averaged spectrum during this observation is not compelling, but if the red variability excess is real it may be indicative of emission from a relatively isolated "hot spot" from a narrow range of radii in the relativistic region of the disk. 

It is interesting to compare our work to that of De Marco et al. (2009; DM09), who have studied the iron K$\alpha$ line variability of a sample of nearby radio-quiet AGN by searching for excess emission in the time-energy plane. Twelve of their observations overlap with rms spectra from our work: MCG-5-23-16(2), NGC 3516(2), NGC 3783, NGC 4051(1), Mrk 766(2), NGC 4593, MCG-6-30-15(1,2), NGC 5548(2), NGC 7314(1), Ark 564 and NGC 7469(2). Their method differs from ours in that it is optimised to search for narrow, transient features in the excess maps, whereas ours will characterize any variability, or lack thereof. Comparing the overlapping sample, eight out of the twelve showed no significant variability across the iron K band due to narrow transient lines in the excess maps. This includes MCG-5-23-16(2), NGC 3516(2) and NGC 4051(1) in which we found variability across the iron line region that must then be due to a variable broad line, rather than narrow transient lines.

DM09 found significant variability in four observations for which we also have rms spectra. NGC 7314(1) showed significant variations in the 5.4-6.1 keV band. Our rms spectrum of this observation was difficult to classify although there is a significant variability excess in the 5.4-5.8 keV bin that could be a redshifted Fe K$\alpha$ component as found in a 2002 Chandra observation (Yaqoob et al. 2003). Both observations of MCG-6-30-15 were found to exhibit Fe K band variability in the excess maps of DM09. This agrees broadly with our result for MCG-6-30-15(1), although we note that the variability across the iron line region in this observation is unlikely to be due solely to narrow line components as the excess extends down to 4.6 keV and it overlaps well with the time-averaged line profile. The rms spectrum of the second MCG-6-30-15 observation shows narrow variability to the blue side. Comparison with the excess maps is more difficult in this case, as DM09 considered each of the three \xmm\ revolutions separately, whereas we have combined them. However, we note that at least one of these does show blue variability, consistent with our result. Finally NGC 3783(2) has a variability excess in the rms spectrum that corresponds to the weak broad and redshifted iron line. DM09 also find significant variability between 5.4-6.1 keV in this source, in agreement with the rms spectrum. Overall our results confirm the work of DM09 in the sense that both studies show that, contrary to some previous reports, variations in the iron K$\alpha$ band are not uncommon. They also confirm that work in the sense that these variations are extremely difficult to interpret.


There is a remarkable diversity in the shapes of the rms variability profiles of the iron line K$\alpha$ lines in Seyfert galaxies. The majority of objects show some form of variability, although this is restricted to the line wings, with the core remaining relatively constant. The latter almost certainly arises from a great distance, such as in the molecular torus. Arguably the most promising explanation for the diversity in the broad line variability is relativistic effects, such as gravitational light bending, which are able to explain a variety of line behaviours. The flexibility of such models, while useful in interpreting our results, greatly reduce their predictive power. Without clearly defined predictions which can be tested with observations, it will be difficult to unambiguously determine the origin of the broad emission in AGN. 

\section{Acknowledgements}

This paper is based on observations taken with the XMM-Newton Satellite, an ESA science mission with contributions by ESA Member States and USA. We thank Iossif Papadakis for his helpful discussions and Paul O'neill for development of the analysis tools that contributed to this work. KN thanks the Royal society for support and SB acknowledges funding from a Science and Technology Facilities Council (STFC) studentship. We also thank the anonymous referee for their useful comments.

\end{document}